# Diffusion Posterior Sampling for Mitigation of Spatially-Variant System Blur in Ultra-High-Resolution CT of Trabecular Bone

T. Ye, G. Shi, A. Sivakumar, F. J. Quevedo Gonzalez, K. F. Moseley, J. A. Carrino, J. H. Siewerdsen, and W. Zbijewski*

***Abstract— Objective:*** **CT spatial resolution degrades with distance from isocenter. We evaluate diffusion posterior sampling (DPS) deblurring for restoring structural detail and harmonizing texture features (TFs) of off-center trabecular bone to isocenter in ultra-high resolution CT (UHR CT).** ***Methods:*** **DPS pairs a diffusion prior, trained on isocenter UHR CT patches of trabecular bone from four human femora, with a likelihood term enforcing consistency with the observed image. Two variants were tested on a held-out femur at radial shifts of 90, 130 and 160 mm: KnownBlurDPS, with an anisotropic Gaussian kernel derived from modulation transfer function measurements, and BlindDPS, which estimated the kernel jointly with the image. Restored patches were compared with isocenter ground truth by structural similarity (SSIM), gradient magnitude similarity deviation (GMSD), and concordance correlation coefficient (CCC) of 40 gray-level co-occurrence and run-length TFs.** ***Results:*** **Both DPS variants improved SSIM and GMSD by 1–10% and mean TF CCC by 30–130%. The number of reproducible TFs (CCC > 0.80) at 90/130/160 mm rose from 13/2/1 without deblurring to 33/35/29 (KnownBlurDPS) and 40/36/16 (BlindDPS). At 90 mm, the accuracy of a linear classifier predicting the shift from TFs dropped from 79% to chance level with BlindDPS.** ***Conclusion:*** **DPS partly restores trabecular detail and TF reproducibility lost to spatially-variant blur in UHR CT; the blind variant matches the measured-kernel variant up to 130 mm.** ***Significance:*** **DPS deblurring can standardize trabecular bone TFs against patient mis-centering, potentially improving the robustness of TF-based CT radiomics; its blind form needs no CT system characterization.**

*Index Terms*—**Computed tomography, diffusion models, image restoration, radiomics, spatial resolution, trabecular bone.**

## I. Introduction

Recent advances in x-ray CT hardware have resulted in the introduction of ultra-high resolution (UHR) capabilities on several new clinical systems, e.g., Canon Aquilion Precision and Siemens Naeotom Alpha. The imaging chains of these devices comprise x-ray sources with small focal spots (e.g., nominally 0.4 × 0.5 mm on the Aquilion Precision) and either highly optimized energy-integrating detectors or novel photon-counting detectors to achieve detector pixel sizes at the isocenter of 0.25 mm (Aquilion Precision) [1], [2] and ~0.2 mm (Naeotom Alpha) [3]. The resulting minimum resolvable detail approaches ~0.25 mm, a ~2× improvement relative to previous CT generations [1], [2], [3]. With such fine detector pixels, source-side effects dominate the spatial resolution of UHR CT, resulting in system blur that is less stationary across the field of view (FOV) than in conventional systems. Specifically, the radial component of spatial resolution in UHR CT degrades with distance from the isocenter primarily because of elongation of the axial-plane projection of the longitudinal (cross-slice) aspect of the x-ray focal spot; the azimuthal component also worsens towards the edge of the FOV due to x-ray source motion during each detector integration period [1].

One of the consequences of nonstationary spatial resolution in CT is that quantitative biomarkers derived from CT images may become irreproducible – i.e., produce inconsistent values for the same underlying object – between different locations in the FOV. This effect has been demonstrated for grayscale texture features (TFs) measured in regions of trabecular bone, which were found to exhibit a loss of reproducibility (reduced concordance correlation coefficient, CCC) with radial shift from isocenter that was more pronounced in UHR than in conventional normal-resolution (NR) CT [4].

Since TFs are a common component of radiomic models for disease detection and outcome prediction, their reproducibility to scanner settings and, as studied here, location in the FOV, is an important consideration for precision medicine. In orthopedics, TFs of trabecular bone have been demonstrated to correlate with micro-CT measures of trabecular morphology [5], [6] and help to enable radiomic discrimination between patients with and without osteoporotic vertebral fractures [7] and estimation of vertebral failure load [8]. To ensure that such predictive applications are not confounded by the effects of

Manuscript is submitted on November 21, 2026. This work was supported by NIH R01-EB-029446.

T. Ye, G. Shi, and A. Sivakumar are with the Department of Biomedical Engineering, Johns Hopkins University, Baltimore, Maryland, USA.

F. J. Quevedo Gonzalez is with the Department of Biomechanics, Hospital for Special Surgery, New York, USA.

K. F. Moseley is with the Department of Endocrinology, Johns Hopkins University, Baltimore, Maryland, USA.

J. A. Carrino is with the Department of Radiology and Imaging, Hospital for Special Surgery, New York, USA.

J. H. Siewerdsen is with the Department of Imaging Physics, The University of Texas MD Anderson Cancer Center, Houston, Texas, USA.

*W. Zbijewski is with the Department of Biomedical Engineering, Johns Hopkins University, Baltimore, Maryland, USA (correspondence e-mail: wzbijewski@jhu.edu).

location of the anatomy of interest in the FOV, methods to harmonize TFs across different radial shifts need to be developed.

Diffusion posterior sampling (DPS) [9] performs image restoration by pairing a pretrained diffusion generative prior with a likelihood term that enforces consistency with the observed measurements. DPS methods avoid hand-crafted priors of traditional restoration techniques and obviate the need for large, paired training datasets of supervised end-to-end networks (DPS may also potentially be better generalizable in out-of-distribution data). Following on earlier work where DPS was applied in deblurring of natural images [9], [10], [11] and CT reconstruction [12], [13], we investigate its feasibility in mitigating the effects of spatially-variant CT blur in UHR imaging of trabecular bone. We assess the performance of DPS using both conventional image fidelity (IF) metrics relative to ground truth (GT) at isocenter, as well as TF reproducibility (CCC).

If DPS-processed images exhibit better reproducibility of trabecular bone TFs throughout the FOV than the original scans, then DPS deblurring can potentially be used for TF standardization in CT datasets where the location of the site of interest has not been precisely controlled. Other approaches for TF harmonization have been proposed. For example, Gang et al. [14] and Yuan et al. [15] removed the effects of known blur and noise on the gray-level co-occurrence matrix (GLCM), the former by deconvolution in the image domain (blur) and in the GLCM domain (noise), the latter with an end-to-end neural network. Sadia et al. [16] improved the reproducibility of GLCM and gray-level run-length matrix (GLRLM) features across reconstruction settings by synthesizing standardized images with a conditional latent diffusion model trained on paired data. In contrast, our method requires neither paired training data nor, in its blind form, prior knowledge of the blur. It produces a deblurred image, which can then be used to extract standardized TFs and perform visual assessment and other downstream tasks.

The evaluation studies below compare two variants of DPS. In KnownBlurDPS, the blur at any radial displacement (the forward operator in the likelihood-maximization step of DPS) is assumed to be known from physical imaging performance measurements (here using data from [1]). Since a complete characterization of spatially-variant CT blur may be impractical, KnownBlurDPS is included in the evaluations primarily as an idealized reference for the second variant of DPS, which performs blind deconvolution (BlindDPS) [17]. While Ref. [17] estimates the unknown kernel through an additional diffusion model, our approach exploits the fact that the major nonstationary aspects of CT resolution can be approximated by an anisotropic Gaussian blur with shift-dependent widths. This low-dimensional representation can be estimated by gradient-based optimization jointly with the image, removing the need for the second diffusion model. Both DPS variants are evaluated for mitigation of nonstationary CT blur and standardization of bone TFs across radial shifts in experimental data of human femora acquired on a Canon Aquilion Precision UHR CT.

## II. Methods

### A. Experiments and Data Acquisition

The imaging data was acquired on a Canon Aquilion Precision CT (Canon Medical Systems, Japan) in the UHR mode, which combines 0.25 mm detector elements, 0.25 mm slice thickness and a 0.4 × 0.5 mm focal spot to achieve a limiting detail size of ~0.25 mm at the isocenter [1], [2], [4]. Experiments involved five human femora. Prior to imaging, the surrounding soft tissues were removed from each femur and six fiducial markers were attached to bone surface to aid subsequent rigid registration steps (Section II-D).

CT scans were obtained separately for each of the five femora placed at the isocenter and for one of the femora (the test bone) at eight additional radial shifts, with distance from the isocenter ranging from 60 mm to 165 mm in 15 mm steps (Figs. 1 and 2). The radial shifts were executed by translating the bone parallel to CT couch surface. For each position, the CT rotation time was set to the slowest setting of 1.5 s, consistent with high-resolution imaging protocols for orthopedic applications; at this scan speed, the azimuthal blur caused by source motion during each detector integration period is minimized but not eliminated. Volumes were reconstructed with the FC30 bone kernel on a 1024 × 1024 matrix over a 129.9 mm field of view, giving a voxel size of 0.127 × 0.127 × 0.25 mm.

### B. DPS Deblurring

#### 1) Unconditional Diffusion Prior

Both DPS variants use the same unconditional denoising diffusion probabilistic model (DDPM) [18] to encode the distribution $p(x_0)$ of CT image patches of trabecular bone ($x_0$) acquired at the isocenter. Briefly, the model is developed by first applying forward diffusion by sequentially adding Gaussian noise over $T$ steps to each clean training image $x_0$. Denoting the variance of the noise added at step $t$ (variance schedule) as $\beta_t$ and defining $\alpha_t = 1 - \beta_t$, $\overline{\alpha}_t = \prod_{s=1}^{t}(1 - \beta_s)$, the conditional distribution of the noisy image $x_t$ given $x_0$ is given by:

$$q(x_t|x_0) = \mathrm{N}\left(x_t;\, \sqrt{\overline{\alpha}_t}x_0, (1 - \overline{\alpha}_t)\mathbf{I}\right), \tag{1}$$

where $\boldsymbol{I}$ is the identity matrix and $\mathrm{N}(x_t;\, \mu, \Sigma)$ is the Gaussian probability density with mean $\mu$ and covariance $\Sigma$ evaluated at $x_t$. A network $\varepsilon_\theta(x_t, t)$ is then trained to predict the added noise by minimizing $\ell_1$ loss

$$\mathrm{L}(\theta) = \mathrm{E}_{t,x_0,\varepsilon}\|\varepsilon - \varepsilon_\theta(x_t, \mathrm{t})\|_1 \tag{2}$$

with $x_t = \sqrt{\overline{\alpha}_t}x_0 + \sqrt{1 - \overline{\alpha}_t}\varepsilon$ and $\varepsilon \sim N(0, \mathbf{I})$. One can then obtain samples from the image patch distribution by reverse sampling the diffusion process: starting from $x_T \sim \mathrm{N}(0, \mathbf{I})$, each step removes the predicted noise and adds back a smaller amount of noise,

$$x_{t-1} = \frac{1}{\sqrt{\alpha_t}}\left(x_t - \frac{\beta_t}{\sqrt{1 - \overline{\alpha}_t}}\varepsilon_\theta(x_t, t)\right) + \sigma_t z,\ \mathrm{z} \sim \mathrm{N}(0, \mathbf{I}) \tag{3}$$

until $x_0$ is reached. Here $\sigma_t^2 = \beta_t(1 - \overline{\alpha}_{t-1})/(1 - \overline{\alpha}_t)$ is the variance of the added noise at step $t$; no noise is added at the final step.

Importantly, the trained denoising network $\varepsilon_\theta(x_t, t)$ provides an estimate of the score (the gradient of the log-density with respect to $x_t$) of $p(x_t)$, the distribution of the noisy images $x_t$ obtained from $p(x_0)$ through the forward process of Eq. (1),

$$\nabla_{x_t} \log p(x_t) \approx -\frac{\varepsilon_\theta(x_t,t)}{\sqrt{1-\bar{\alpha}_t}}. \quad (4)$$

The training data and settings are described in Section II-C.

*2) Posterior Sampling for Deblurring With a Known Kernel (KnownBlurDPS)*

DPS [9] adapts the unconditional sampling process to draw samples from the posterior $p(x|y)$, where $y$ is in our case the measured image corrupted by radial shift blur and $x$ is the unknown trabecular patch at the isocenter, i.e., a sample of $p(x_0)$; the subscript 0 is omitted for brevity. To this end, Bayes' rule is applied at every reverse step $t$ to compute the score of the posterior from the score of the forward-propagated diffusion prior ($\nabla_{x_t} \log p(x_t)$, Eq. (4)) and a likelihood model $p(y|x)$:

$$\nabla_{x_t} \log p(x_t|y) = \nabla_{x_t} \log p(x_t) + \nabla_{x_t} \log p(y|x_t). \quad (5)$$

Because $p(y|x_t)$ is intractable, DPS approximates it by $p\big(y|\hat{x}_0(x_t)\big)$, i.e., the likelihood at the estimate of the clean image available at step $t$ (the Tweedie estimate):

$$\hat{x}_0(x_t) = \frac{x_t - \sqrt{1-\bar{\alpha}_t}\varepsilon_\theta(x_t,t)}{\sqrt{\bar{\alpha}_t}}. \quad (6)$$

We use the following forward model to relate the latent image patch at isocenter $x$, the radial shift blur kernel at distance $r$ from the isocenter $k(r)$, and the measured image patch at $r$ (denoted $y(r)$):

$$y(r) = k(r) * x + n(r), \quad (7)$$

where * denotes convolution and $n(r)$ is zero-mean Gaussian noise at $r$ with variance $\sigma_n(r)^2$. For simplicity, we assume that the noise is uniform within the small patch $y(r)$ and uncorrelated, thus ignoring the effects of the reconstruction kernel. The resulting likelihood is:

$$p(y|x) \propto \exp\left(-\frac{\|y - k * x\|_2^2}{2\sigma_n{}^2}\right), \quad (8)$$

where we omitted $r$ for clarity. Substituting Eqs. (6)–(8) into Eq. (5) and assuming that $k$ is known (Section II-B3), each reverse step of KnownBlurDPS consists of the unconditional update of Eq. (3), giving an intermediate sample $x'_{t-1}$, followed by a data-consistency correction,

$$x_{t-1} \leftarrow x'_{t-1} - \alpha\big(\nabla_{\hat{x}_0} \|y - k * \hat{x}_0(x_t)\|_2\big), \quad (9)$$

where the use of the $\ell_2$ norm of the residual $y - k * \hat{x}_0$ follows [9]. The gradient is taken with respect to $\hat{x}_0$, without backpropagation through the denoising network, following [13]. The constant factor $\frac{1}{2\sigma_n^2}$ of Eq. (8) is absorbed into the step size $\alpha$. In our implementation, the gradient step of Eq. (9) was taken with the Adam optimizer [19] instead of plain gradient descent. Adam was used due to its generally advantageous stability and convergence rate. One Adam update was performed per reverse step, with the moment estimates carried over between steps (decay rates in Section II-D).

Following [13], we adopted a jump start in which sampling begins at an intermediate step $t_0 \ll$ T from a forward-diffused version of the input blurred patch $y$, $x_{t_0} \sim \mathrm{q}\big(x_{t_0}|y\big)$ (Eq. (1) with $x_0$ replaced by $y$), and only the last $t_0$ reverse steps are performed. In CT reconstruction using DPS, it has been shown that jump start reduces both the number of sampling steps and the variability of the sampled images by anchoring their low-frequency content to measured data $y$ [13].

*3) Known Kernel Settings for Shift-Variant Blur in UHR CT*

KnownBlurDPS assumes that $k$ is a bivariate Gaussian blur kernel with location-dependent width:

$$k_\omega(u,v,r) = \frac{1}{2\pi\omega_v(r)\omega_u(r)} \exp\left(-\frac{u^2}{2\omega_u(r)^2} - \frac{v^2}{2\omega_v(r)^2}\right), \quad (10)$$

where $u$ and $v$ are image slice coordinates (mm), and $\omega(r) = (\omega_u(r), \omega_v(r))$ are the corresponding characteristic widths of the kernel (mm). The image coordinates are defined as follows: $u$ is parallel to CT couch surface and thus parallel to the translation direction of the test femur; $v$ is a perpendicular axis pointing up from the couch surface. In our experimental setup, where the bones were placed directly on the couch with minimal $v$ elevation, $\omega_u(r)$ is associated with the radial component of system blur and $\omega_v(r)$ is associated with the azimuthal component.

The parameters of $k_\omega(u,v,r)$ were obtained from 1D radial and azimuthal modulation transfer functions (MTFs) reported for gantry rotation time of 1.5 s in [1]. Specifically, the radial and azimuthal MTFs measured at shift distances of 75, 110, 145, and 185 mm were each normalized by their corresponding MTFs at 40 mm shift (the reported position closest to the isocenter). Next, 1D Gaussians were fit to these normalized MTFs, each given by $\mathrm{MTF}(f) = \exp(-2\pi^2\omega^2 f^2)$, where $f$ was the spatial frequency along the corresponding (radial or azimuthal) direction. Since this form of the MTF corresponds to a 1D Gaussian PSF with characteristic width $\omega$ (mm), the fits yielded the spatial-domain widths $\omega_u(r)$ and $\omega_v(r)$ in $k_\omega(u,v,r)$ directly (Table I).

TABLE I
RADIAL AND AZIMUTHAL FREQUENCIES AT 10% MTF ($f_{10}$) AT SELECTED SHIFT DISTANCES EXTRACTED FROM THE MEASUREMENTS IN [1], TOGETHER WITH THE CHARACTERISTIC WIDTHS OF THE CORRESPONDING GAUSSIAN KERNELS. THE RADIAL WIDTH AT 185 MM WAS NOT USED FOR KERNEL ESTIMATION (SEE TEXT).

| Radial shift (mm) | Radial $f_{10}$ (lp/mm) | Azi $f_{10}$ (lp/mm) | $\omega_u$ (mm) | $\omega_v$ (mm) |
|---|---|---|---|---|
| 40 | 1.99 | 2.00 | NA | NA |
| 75 | 1.95 | 1.87 | 0.064 | 0.076 |
| 110 | 1.85 | 1.67 | 0.102 | 0.124 |
| 145 | 1.68 | 1.50 | 0.135 | 0.157 |
| 185 | 1.16 | 1.31 | 0.280 | 0.191 |

For radial shifts that were not explicitly characterized in [1], the characteristic widths were obtained by linear interpolation between the two nearest shift distances in Table I. The only exception was the radial width at 160 mm. The radial width measured at 185 mm departed from the linear trend of the values at 75–145 mm and was not used. The radial width at 160 mm was therefore obtained by extrapolating a straight line fitted by least squares to the radial widths at 75, 110, and 145 mm. The resulting radial and azimuthal widths were 0.080 and 0.097 mm at 90 mm, 0.121 and 0.143 mm at 130 mm, and 0.151 and 0.170 mm at 160 mm. The kernels were then discretized on a 31 × 31-pixel grid and normalized to unit sum.

*4) Posterior Sampling With an Unknown Kernel (BlindDPS)*

In BlindDPS, the blur is unknown and estimated jointly with the image. In the original implementation [17], the DPS image prior was paired with a second diffusion prior over blur kernels. We use a simpler variant: the kernel is restricted to a two-parameter Gaussian as in Eq. (10) and $\omega(r) = (\omega_u(r), \omega_v(r))$ is optimized directly. This is done separately for each radial shift $r$; we therefore drop $r$ from the subsequent derivations.

At each reverse step, after the image update of Eq. (9), $\omega$ is updated using the gradient of the data-consistency term with respect to $\omega$ ($\nabla_\omega$) obtained by automatic differentiation:

$$\omega \leftarrow \omega - \alpha_k(\nabla_\omega \|y - k_\omega * \hat{x}_0(x_t)\|_2), \tag{11}$$

where $\alpha_k$ is the kernel step size; as in Eq. (9), this step was taken with a separate Adam optimizer with its own moment estimates.

## C. *Diffusion Prior Training*

**Training data.** The same diffusion prior was used by both DPS variants. The prior distribution represented 2D trabecular bone images with minimal spatially-variant blur, obtained by training on axial slices of the isocenter scans of the four femora that were not used for testing.

For each training femur, non-overlapping image patches of 128 × 128 pixels (16.2 mm × 16.2 mm) were extracted from the proximal (femoral head and neck) and distal (condylar) 20% of the bone, where trabecular bone is most abundant. Every other axial slice within these regions was used to limit redundancy between adjacent 0.25 mm slices.

In each slice, candidate patches were first seeded on a regular grid within a bone outline given by the largest connected contour determined by intensity thresholding the image. A patch was accepted for inclusion in training data only if its four corners and four edge midpoints were inside the bone contour, its center was at least 64 pixels from all previously accepted patches in both $u$ and $v$, its maximum intensity was below 1500 HU (no cortical bone) and its mean intensity was above 150 HU (not dominated by soft tissue). In total, 10,594 2D axial patches were obtained from the four training femora.

**Model design and training procedure.** The noise-prediction (denoising) network $\varepsilon_\theta(x_t, t)$ was a U-Net with 64 base channels, channel multipliers (1, 2, 4, 8, 16), one residual block per resolution level, self-attention at the 16 × 16 resolution, group normalization (16 groups) and dropout of 0.2, operating on 128 × 128 single-channel patches whose intensities were clipped to the [−1000, 1700] HU window and linearly mapped to [−1, 1]. The forward process used $T = 1500$ steps with a cosine variance schedule [20]. The network was trained for 100,000 iterations with Adam (learning rate $4 \times 10^{-5}$, batch size 64) using the $\ell_1$ noise-prediction loss of Eq. (2); an exponential moving average of the weights (decay 0.995) was used for sampling. Training was performed on two NVIDIA RTX A6000 GPUs.

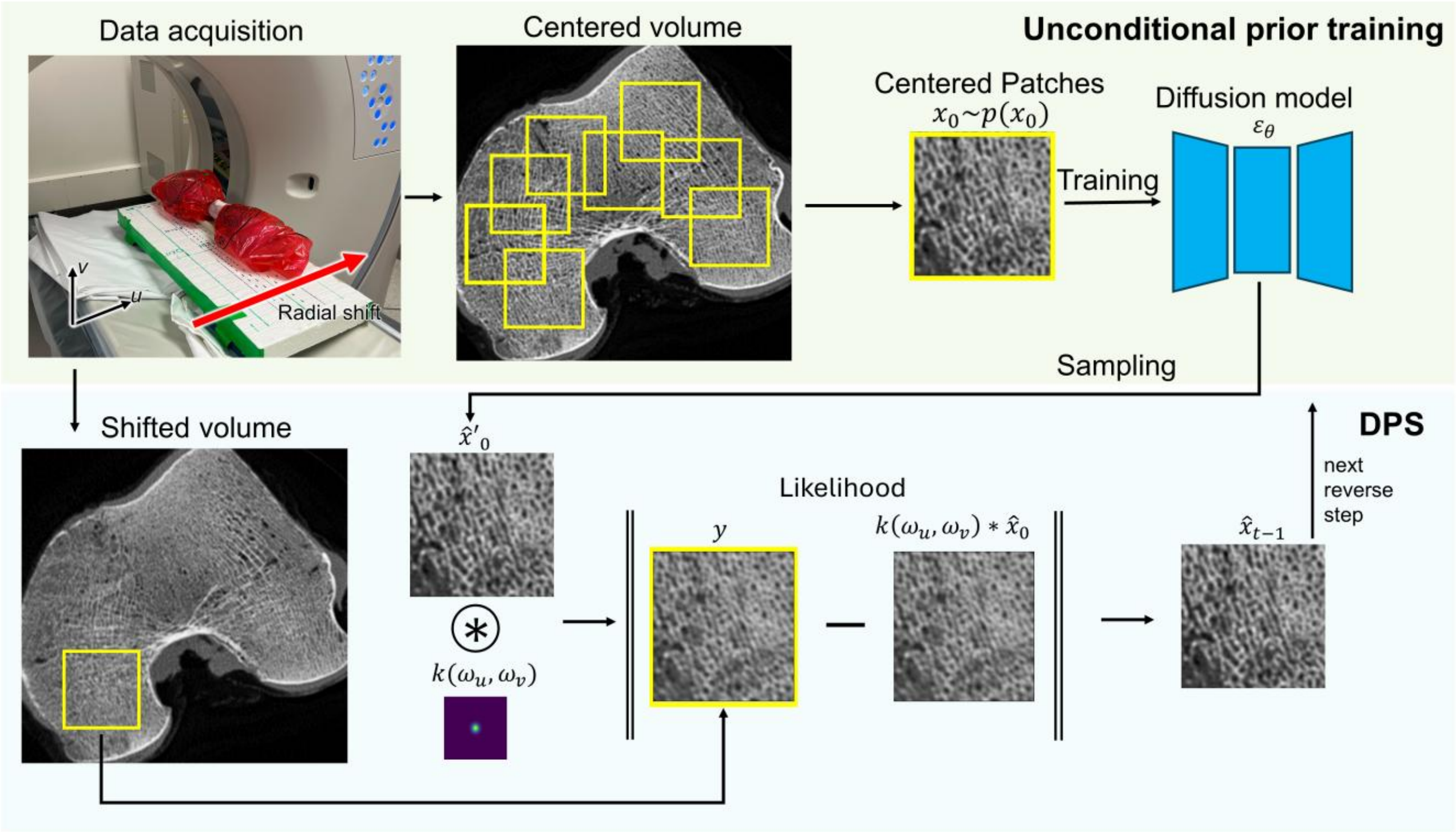

Fig. 1. Workflow of DPS training, deployment, and validation. An unconditional diffusion model representing trabecular bone patches imaged at UHR CT isocenter was trained on scans of four centered femora. DPS was then applied to restore image patches extracted from UHR CTs of a test femur placed at different radial shifts. A likelihood update term was incorporated in diffusion model sampling to guide the generation of images from the target (isocenter) distribution toward the posterior given the observed shifted patch. The blur parameters (dependent on radial shift) were either extracted from resolution measurements (KnownBlurDPS) or estimated jointly with the patch likelihood update (BlindDPS).

## D. *Evaluation Experiments*

**Evaluation workflow (Fig. 2).** A total of 215 GT patches whose centers fell within 0–15 mm of the isocenter (to ensure that GT was minimally affected by spatially-varying blur) were sampled in the isocenter scan of the test femur using the same procedure as for the training data. Three of the eight shifted scans were used for evaluation, selected to span moderate to large radial displacements. For each selected scan, a six-degree-of-freedom (6-DoF) rigid transformation between this volume (moving image) and the centered scan (fixed image) was estimated by fiducial marker registration. The inverse of the transformation was applied to GT patch coordinates and test patches were obtained by resampling the shifted volumes at these transported coordinates with linear interpolation. The mean radial distances of the resulting test patches in the three shifted scans were 91, 131 and 162 mm; the three groups are referred to by their mean displacements rounded to 90, 130 and 160 mm. We denote these input datasets as NoDeblur below. For each shift group, three of the 215 patches were reserved for hyperparameter selection and 212 were used for the final evaluation.

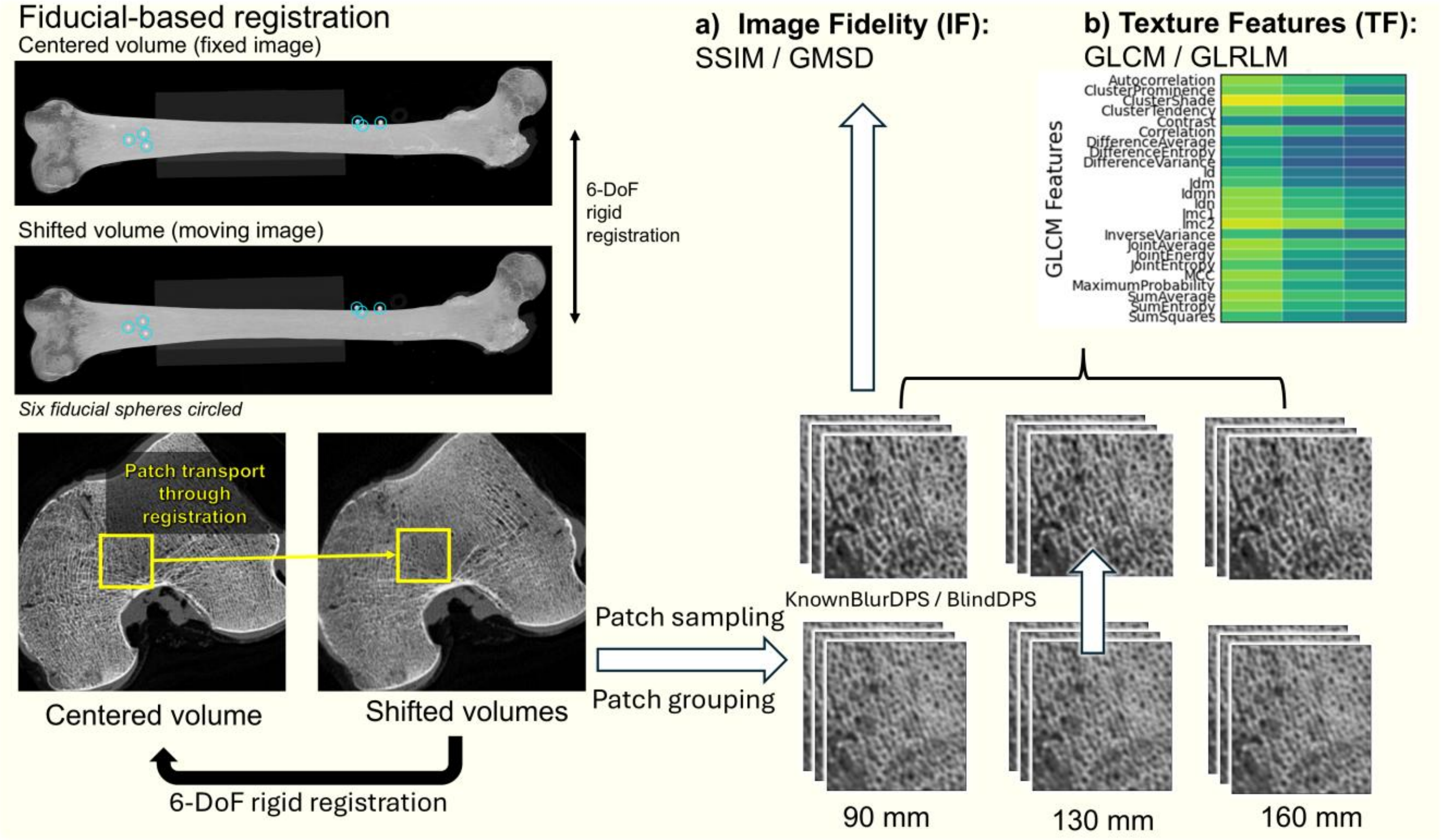

Fig. 2. Evaluation workflow. Test patches were seeded in the centered volume of a held-out test femur and transported to its shifted scans by fiducial-based rigid registration (fiducials circled in maximum-intensity projections). After deblurring with KnownBlurDPS and BlindDPS, the original and processed shifted patches were grouped by their mean radial displacement and compared to the ground-truth (centered) patches by (a) IF metrics (SSIM, GMSD) and (b) TF concordance (CCC).

**DPS settings.** Sampling hyperparameters were selected on the reserved patches by grid search to maximize the mean structural similarity index measure (SSIM) against GT.

For KnownBlurDPS, the resulting jump-start step was $t_0$ = 35 (searched values: {15, 25, 35, 50, 75, 100}) and the image update step size was $\alpha = 4 \times 10^{-3}$ (searched values: {$6 \times 10^{-4}$, $8 \times 10^{-4}$, $1 \times 10^{-3}$, $2 \times 10^{-3}$, $4 \times 10^{-3}$, $6 \times 10^{-3}$, $8 \times 10^{-3}$}). For each shift group, the characteristic radial and azimuthal widths of the known kernel were set to values estimated from [1] at the nominal mean displacement of that group.

For BlindDPS, $t_0$ was fixed at 100 to ensure sufficient number of kernel-update steps. The image step size $\alpha$ was optimized at $2 \times 10^{-3}$ (searched values: $\{1, 2, 4, 6, 8\} \times 10^{-3}$) for all evaluated kernel step sizes $\alpha_k$. Among the kernel step sizes (searched values: {0.05, 0.1, 0.5, 1}), $\alpha_k = 0.5$ reached a marginally higher mean SSIM than the second-best setting of $\alpha_k = 0.1$ (0.852 vs. 0.849, well within the patch-to-patch SSIM standard deviation of 0.06). However, $\alpha_k = 0.1$ produced kernels that varied smoothly with shift distance, whereas the estimates obtained with $\alpha_k = 0.5$ were unstable both on the reserved patches and in a subsequent run on the test patches; the former setting was therefore selected. The kernel was initialized at $\omega_u = \omega_v = 0.2$ mm, held fixed for the first 40 steps and updated during the remaining 60 steps. The Adam decay rates were $\beta_1 = 0.9$ and $\beta_2 = 0.999$ for both the image updates (KnownBlurDPS and BlindDPS) and the kernel updates (BlindDPS).

*1) IF Metrics: SSIM and GMSD*

IF evaluations of deblurring effectiveness involved two metrics, SSIM and gradient magnitude similarity deviation (GMSD) [21].

SSIM compares local means $\mu$, variances $\sigma^2$ and covariance $\sigma_{dg}$ of the shifted patch $d$ and the corresponding GT patch $g$ within a sliding window (here, 7 × 7 pixels) and averages the result over the patch,

$$\mathrm{SSIM}(d, g) = \frac{(2\mu_d\mu_g + C_1)(2\sigma_{dg} + C_2)}{(\mu_d{}^2 + \mu_g{}^2 + C_1)(\sigma_d{}^2 + \sigma_g{}^2 + C_2)}, \qquad (12)$$

where $C_1$ and $C_2$ are small stabilizing constants. SSIM varies from −1 to 1 (SSIM = 1 for identical images). Data range for SSIM calculation was set to the joint intensity range of each ($d$, $g$) pair.

GMSD compares the gradient magnitudes $m_d$ and $m_g$ (Sobel operator) of the two patches in terms of the standard deviation of pixel-wise gradient magnitude similarity (GMS):

$$\mathrm{GMS}(i) = \frac{2m_d(i)m_g(i) + \mathrm{c}}{m_d^2(i) + m_g^2(i) + \mathrm{c}},$$

$$\mathrm{GMSD} = \sqrt{\frac{1}{N}\sum_{i=1}^{N}\left(\mathrm{GMS}(i) - \overline{\mathrm{GMS}}\right)^2}, \quad (13)$$

where $i$ is pixel index, $N$ is number of pixels, $c$ is a stabilizing constant, and $\overline{\mathrm{GMS}}$ is the mean GMS; lower GMSD indicates higher fidelity, with 0 for identical images. Compared to SSIM, GMSD aligns more closely with human perceptual quality [21]. Further, in super-resolution of trabecular bone, gradient-based metrics, but not SSIM, correlated with the accuracy of the recovered microstructural metrics (bone volume fraction, trabecular thickness, and trabecular number) [22].

For each IF metric, paired comparisons between conditions (e.g., blurred vs. deblurred) were performed on the per-patch metric values. The normality of the paired differences was first assessed using the Shapiro–Wilk test; a paired *t*-test was applied when the normality of the differences was not rejected (Shapiro–Wilk $p > 0.05$), and the nonparametric Wilcoxon signed-rank test otherwise. Significance was defined at $p < 0.05$.

*2) TF Extraction and Reproducibility*

GLCM and GLRLM TFs were extracted from the central 2D regions of interest (ROIs, 6 mm × 6 mm) within each test patch in the GT and in shifted volumes without DPS deblurring (NoDeblur) and with deblurring (KnownBlurDPS and BlindDPS). The reproducibility of each TF for each radial displacement group and processing method (NoDeblur vs. DPS deblurring) was quantified by Lin's CCC [23] between TF values in the shifted set, *D*, and in the GT set, *G*:

$$\mathrm{CCC} = \frac{2\sigma_{D,G}}{\sigma_D^2 + \sigma_G^2 + (\mu_D - \mu_G)^2}, \quad (14)$$

where $\mu$ and $\sigma^2$ were the mean and variance of the feature over the 212 test patches for each condition, and $\sigma_{D,G}$ was the covariance between the two conditions. A TF was considered reproducible when CCC > 0.80 (CCC = 1 indicates perfect agreement with the value at isocenter).

Differences in CCC between conditions (KnownBlurDPS vs. NoDeblur, BlindDPS vs. NoDeblur and BlindDPS vs. KnownBlurDPS) were tested for each TF with a paired permutation test [24]. The two conditions were swapped at random for each patch, the CCC difference was recomputed, and the two-sided p-value was obtained from 10,000 such permutations. The p-values were corrected for multiple comparisons separately within GLCM and GLRLM features using the Benjamini–Hochberg procedure [25] to control the false discovery rate (FDR). A TF was considered significantly different when the corrected p-value (q) was below 0.05.

*3) Prediction of Radial Displacement From Image Patch TFs*

Similar to the experiments reported in [4], we evaluated whether the patches processed with either variant of DPS deblurring retain information about their radial displacement. For each shift group, a linear support-vector machine (SVM) was trained on TFs to classify image patches between (i) the NoDeblur dataset at that shift and the isocenter, and (ii) the DPS-restored datasets at that shift and the isocenter. Accuracy was estimated by five-fold cross-validation. The folds were formed by grouping the 212 test patch locations by their slice index ($z$), so that locations belonging to neighboring slices were never simultaneously in training and test data. This approach guards against possible biases due to inter-slice TF correlations. Accuracies are reported as the mean and standard deviation over the five folds.

The cross-validation experiments were repeated with the class labels of all patches (training and test) randomly permuted within each fold (200 permutations) [26] to obtain a distribution of chance-level classification accuracies. Each permuted run was scored by the same statistic as the real one (mean accuracy over the five folds). The resulting distributions were summarized by their median and 2.5th–97.5th percentile range as a baseline of expected SVM performance if the restored TFs were indistinguishable from those of the isocenter.

## III. Results

### A. Visual Assessment

Fig. 3 shows representative NoDeblur patches for the 90, 130 and 160 mm displacement groups (rows) together with their KnownBlurDPS- and BlindDPS-restored versions and the corresponding GT (isocenter) images. The SSIMs of the patches selected for Fig. 3 were close to the median SSIM of their shift group before the deblurring step and were also close to the median SSIM of their DPS-processed shift group after deblurring (within 0.01). Progressive worsening of the spatially-variant system resolution with radial displacement is apparent by comparing trabecular features between NoDeblur and isocenter images in each group (leftmost column vs. rightmost column). Relative to NoDeblur, both DPS deblurring approaches result in sharper delineation of the trabeculae, approaching that of GT for all shifts. However, while the sharpening by either DPS method is more pronounced at large radial displacement, the ability to faithfully reproduce the cancellous structures visible in GT generally worsens with increasing shift. Visually, there is little difference between the outputs of KnownBlurDPS and BlindDPS.

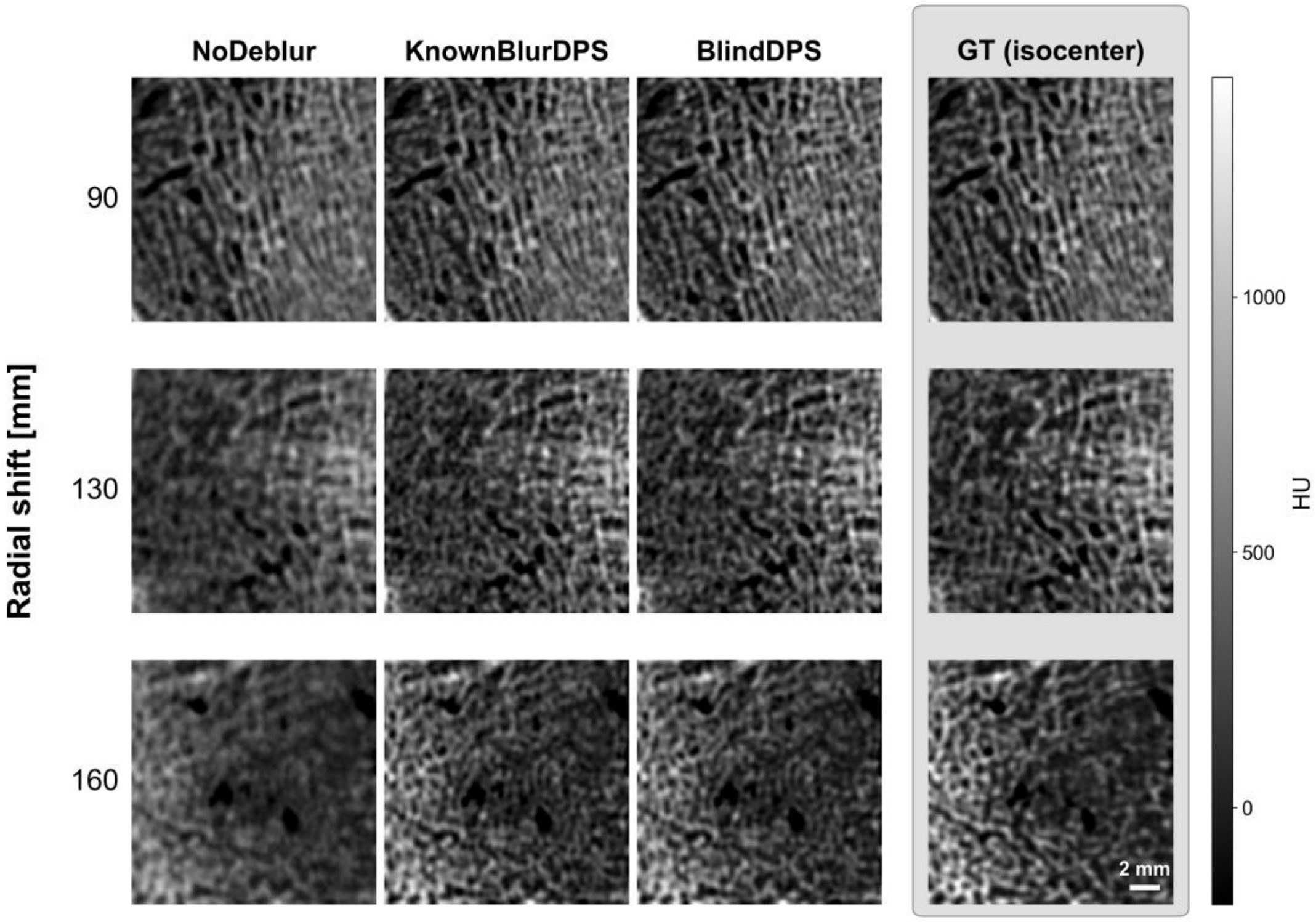

Fig. 3. Representative NoDeblur image patches for radial shifts of 90, 130 and 160 mm (rows) alongside the corresponding KnownBlurDPS and BlindDPS results and the GT images at the isocenter (gray box). Scale bar: 2 mm; gray scale in HU. At each shift, the SSIMs against GT of the representative NoDeblur patch and its restored versions all approach the median SSIM of the shift group for these conditions.

## B. *Image Fidelity Evaluation*

Both KnownBlurDPS and BlindDPS consistently improved SSIM and GMSD relative to NoDeblur across the investigated shift groups, with only minor differences between the two DPS variants (Fig. 4). Consistent with the trends observed in Fig. 3, the relative benefit of DPS is more substantial at large displacements (130 and 160 mm) although the IF metrics of the restored images remain worse than those for NoDeblur at the shortest shift of 90 mm.

For SSIM, median values increased from 0.887/0.795/0.702 (NoDeblur) to 0.901/0.838/0.759 (KnownBlurDPS) and 0.901/0.842/0.754 (BlindDPS), corresponding to improvements of 0.013–0.057 for both methods (all significant, $p < 0.05$). Differences between KnownBlurDPS and BlindDPS were small, with no significant difference at 90 mm ($p = 0.55$) and marginal but statistically significant differences at 130 mm (BlindDPS higher, $p < 0.05$) and 160 mm (KnownBlurDPS higher, $p < 0.05$).

For GMSD (where lower values indicate better IF), median values decreased from 0.152/0.188/0.215 (NoDeblur) to 0.149/0.174/0.192 (KnownBlurDPS) and 0.150/0.173/0.194 (BlindDPS), corresponding to reductions of 0.002–0.022 (all significant, $p < 0.05$). The differences between KnownBlurDPS and BlindDPS were small but significant at all three shifts ($p < 0.05$), with KnownBlurDPS lower at 90 and 160 mm and BlindDPS lower at 130 mm, the same trend as for SSIM at 130 and 160 mm.

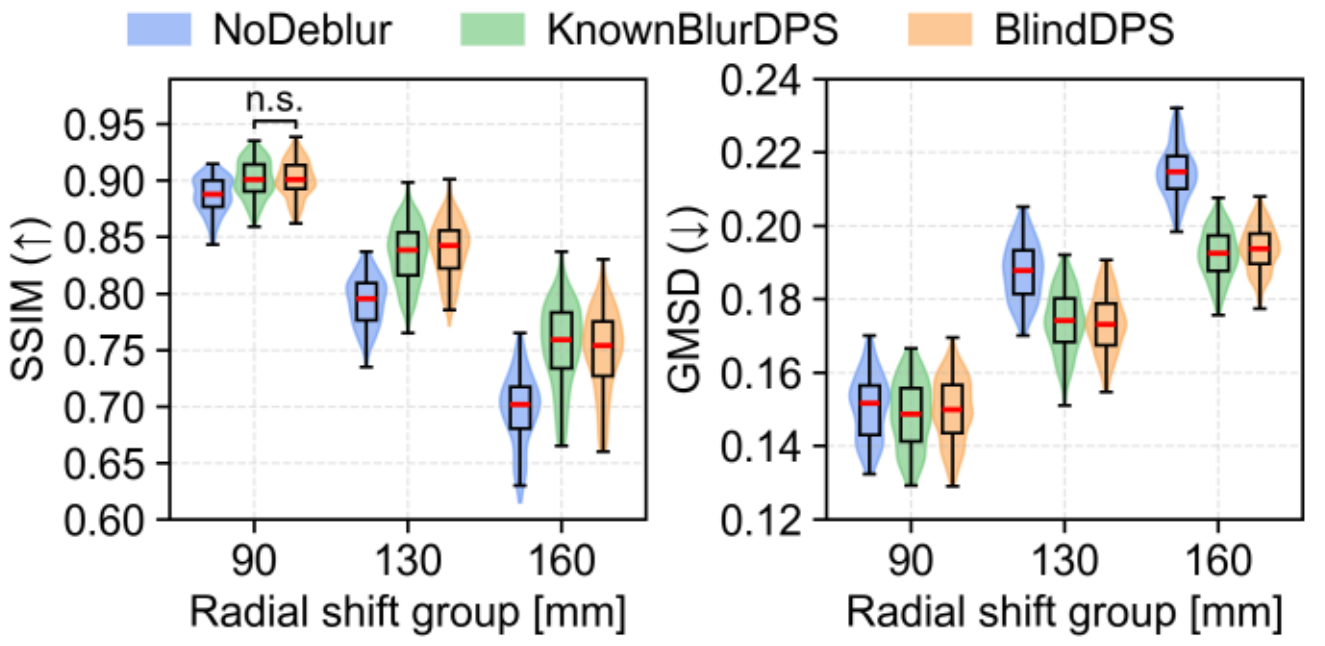

Fig. 4. SSIM and GMSD of NoDeblur, KnownBlurDPS and BlindDPS versus GT for radial shift groups of 90, 130 and 160 mm. The only pairwise difference within a shift group that was not significant (BlindDPS vs. KnownBlurDPS in SSIM at 90 mm, $p = 0.55$) is marked "n.s."; all other pairwise differences were significant ($p < 0.05$).

## C. *Reproducibility of GLCM/GLRLM Texture Features*

Fig. 5 shows CCCs to isocenter of all examined TFs for the NoDeblur and DPS-restored images across radial shift groups. In general, the number of reproducible texture features (CCC > 0.80) in NoDeblur patches decreases with radial distance, consistent with increasing spatially-variant system blur. Both KnownBlurDPS and BlindDPS improve CCC and the number of reproducible features compared to NoDeblur across all radial displacement groups. The number of reproducible features (CCC > 0.80) out of forty total increases from 13/2/1 (NoDeblur) to 33/35/29 (KnownBlurDPS) and 40/36/16 (BlindDPS) at 90/130/160 mm.

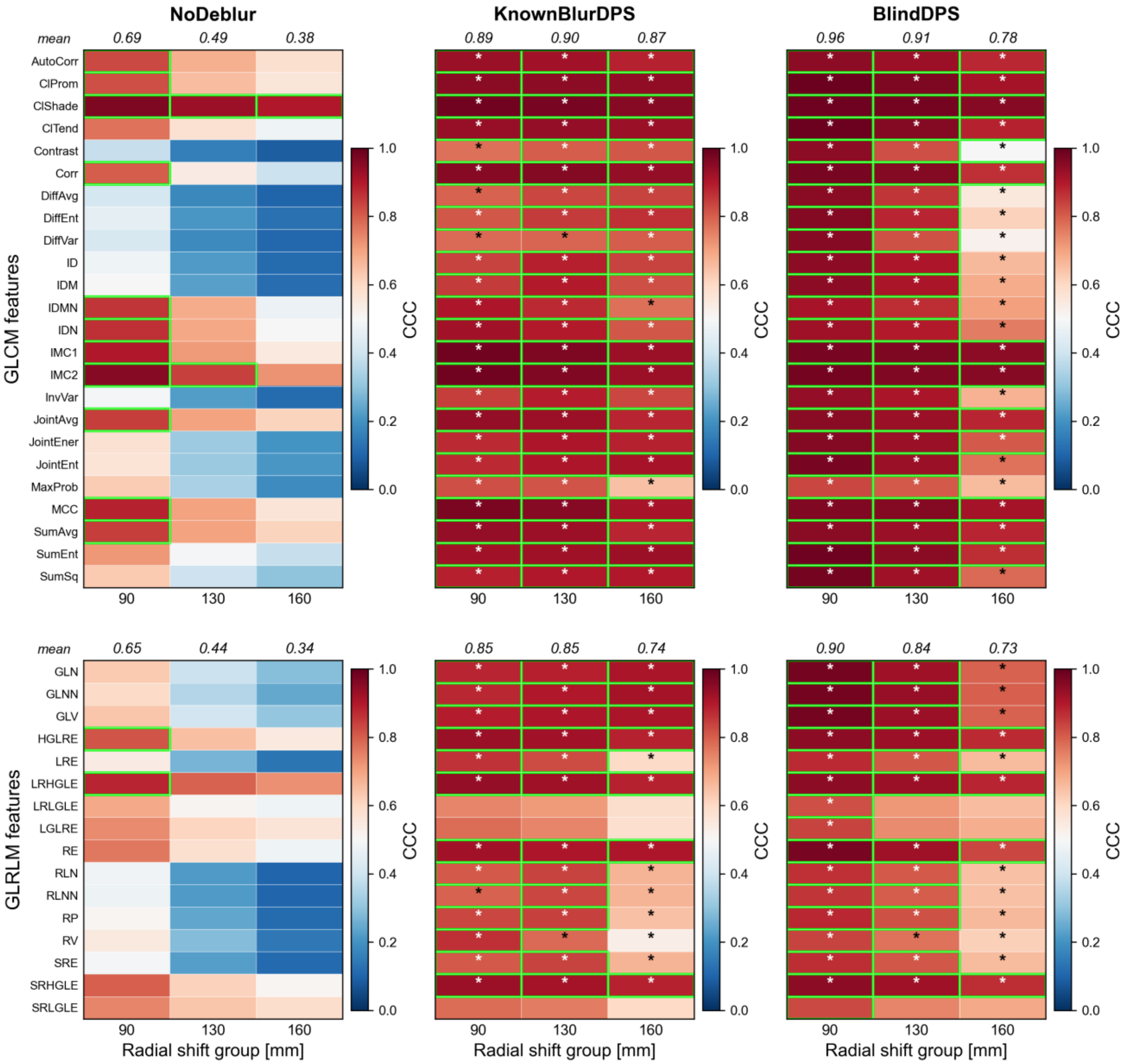

Fig. 5. CCC maps of GLCM and GLRLM texture features versus ground truth, by shift group (90, 130, 160 mm), for NoDeblur, KnownBlurDPS, and BlindDPS. Green outlines mark CCC > 0.80; the number above each column is the mean CCC across the features; stars mark features whose CCC improved significantly over NoDeblur (paired permutation test, Benjamini–Hochberg q < 0.05). Feature acronyms for GLCM features: AutoCorr = Autocorrelation, ClProm = Cluster Prominence, ClShade = Cluster Shade, ClTend = Cluster Tendency, Corr = Correlation, DiffAvg = Difference Average, DiffEnt = Difference Entropy, DiffVar = Difference Variance, ID = Inverse Difference, IDM = Inverse Difference Moment, IDMN = Inverse Difference Moment Normalized, IDN = Inverse Difference Normalized, IMC1 = Informational Measure of Correlation 1, IMC2 = Informational Measure of Correlation 2, InvVar = Inverse Variance, JointAvg = Joint Average, JointEner = Joint Energy, JointEnt = Joint Entropy, MaxProb = Maximum Probability, MCC = Maximal Correlation Coefficient, SumAvg = Sum Average, SumEnt = Sum Entropy, SumSq = Sum Squares. Feature acronyms for GLRLM features: GLN = Gray Level Non-Uniformity, GLNN = Gray Level Non-Uniformity Normalized, GLV = Gray Level Variance, HGLRE = High Gray Level Run Emphasis, LRE = Long Run Emphasis, LRHGLE = Long Run High Gray Level Emphasis, LRLGLE = Long Run Low Gray Level Emphasis, LGLRE = Low Gray Level Run Emphasis, RE = Run Entropy, RLN = Run Length Non-Uniformity, RLNN = Run Length Non-Uniformity Normalized, RP = Run Percentage, RV = Run Variance, SRE = Short Run Emphasis, SRHGLE = Short Run High Gray Level Emphasis, SRLGLE = Short Run Low Gray Level Emphasis.

The mean CCC across the features, given above each column of Fig. 5, confirms that both KnownBlurDPS and BlindDPS achieve substantially better CCC to isocenter than NoDeblur, with most pronounced gains at 130 and 160 mm (mean CCC difference vs. NoDeblur at 90/130/160 mm: +0.20/+0.42/+0.49 for GLCM and +0.21/+0.41/+0.40 for GLRLM with KnownBlurDPS; +0.26/+0.43/+0.40 and +0.26/+0.41/+0.39 with BlindDPS). The results of permutation tests with FDR correction show that all 24 GLCM features improved significantly with both DPS methods at every shift (marked with stars in Fig. 5). Among the 16 GLRLM features, KnownBlurDPS achieved statistically significant gains in CCC for 13/13/13 TFs at 90/130/160 mm. For BlindDPS, significant improvement was found for 15/13/13 TFs at 90/130/160 mm; no TF degraded significantly with either method.

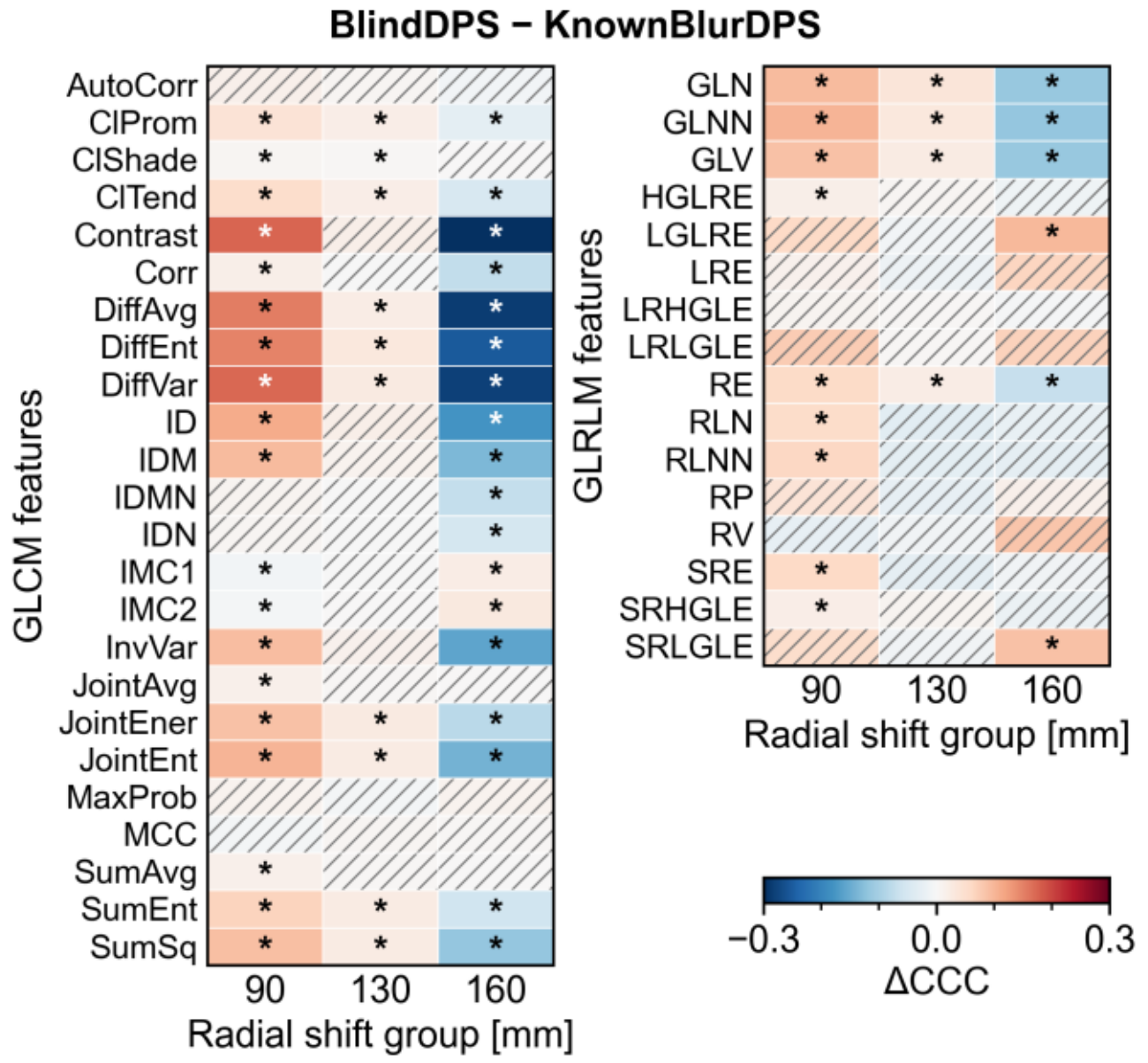

Fig. 6. Differences in CCC (ΔCCC) between BlindDPS and KnownBlurDPS for each GLCM (left) and GLRLM (right) feature and shift group. Color encodes the effect size (red: higher CCC with BlindDPS; blue: higher CCC with KnownBlurDPS; white: no difference). Asterisks mark differences that are significant in the paired permutation test after Benjamini–Hochberg correction ($q < 0.05$); hatched cells are not significant. With 212 paired patches, even small differences can reach significance. Feature acronyms as in Fig. 5.

Fig. 6 summarizes the CCC differences (ΔCCC) between KnownBlurDPS and BlindDPS. For GLCM features at 90 mm shift, 19 of the 24 TFs differ significantly between the two DPS methods, 17 of them in favor of BlindDPS; similarly at 130 mm, 10 features differ significantly, all in favor of BlindDPS. In contrast, KnownBlurDPS outperforms BlindDPS at 160 mm, achieving better CCC for 16 out of the 18 GLCM TFs that exhibited significant differences. At 90 mm, the advantage of BlindDPS is largest for contrast-type features (Contrast, DiffAvg, DiffVar, DiffEnt and ID; ΔCCC +0.11 to +0.17). It is the same group of features (together with IDM and InvVar) that favors KnownBlurDPS at 160 mm (ΔCCC between −0.14 and −0.31).

For GLRLM features, the two approaches differ significantly for 9/4/6 of the 16 features at 90/130/160 mm, in favor of BlindDPS at 90 and 130 mm; at 160 mm, four of the six favor KnownBlurDPS.

As evidenced by mean CCC values in Fig. 5 and ΔCCCs in Fig. 6, the performance of BlindDPS relative to KnownBlurDPS worsens at the largest shift (160 mm), especially for GLCM TFs. To better understand the differences between the two DPS variants, Fig. 7 compares the widths of blur kernels obtained by BlindDPS (distribution over all test patches) to their reference values derived from MTF measurements and used by KnownBlurDPS. BlindDPS kernels agree reasonably well with the experimental data but tend to be slightly wider at 90 mm (within ~20% of in-plane voxel size: median $\omega_u = 0.106$ vs. 0.080 mm and $\omega_v = 0.122$ vs. 0.097 mm) and at 130 mm (0.125 vs. 0.121 mm and 0.152 vs. 0.143 mm). At 160 mm, however, BlindDPS produces kernel estimates that are narrower than the MTF-derived value in the radial direction. As a result, the BlindDPS-estimated blur becomes more anisotropic (median $\omega_v/\omega_u = 1.36$) than the MTF-derived blur ($\omega_v/\omega_u = 1.13$).

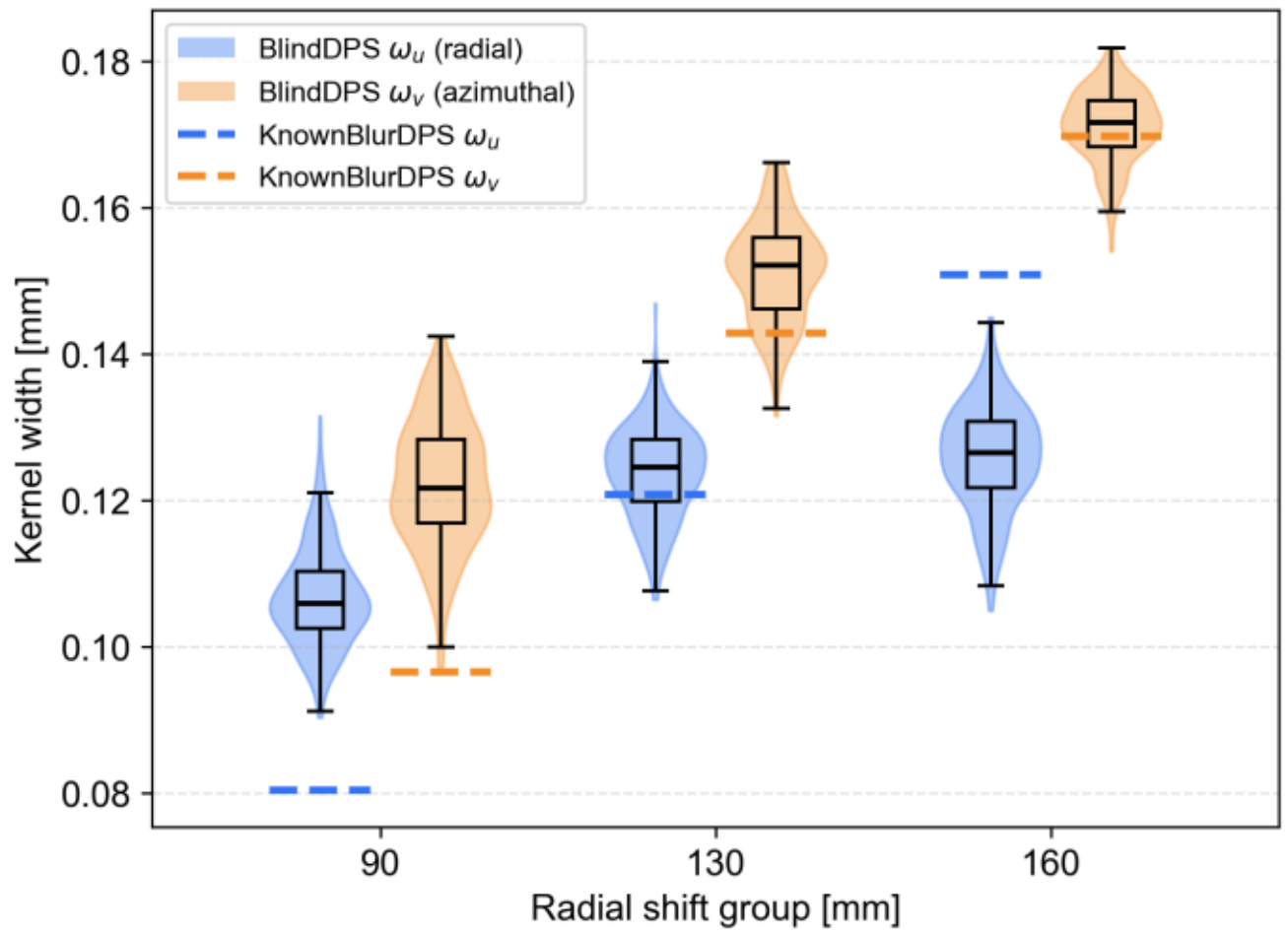

Fig. 7. Distributions of blur kernel widths estimated by BlindDPS for the test patches in each shift group (violin and box plots; radial $\omega_u$ in blue, azimuthal $\omega_v$ in orange) compared with the measurement-derived kernel parameters used by KnownBlurDPS (dashed lines).

### D. *Radial Shift Classification*

Fig. 8 summarizes the classification experiment. Note that in this setting, lower classification accuracy (reduced separability between radial shifts) is generally desirable, as it indicates that the information about displacement within FOV was partially removed from the TFs. The NoDeblur patches are separable from their isocenter counterparts with mean accuracies of 0.79/0.91/0.97 at 90/130/160 mm. Both DPS deblurring methods reduce the separability; the resulting accuracies are 0.51/0.67/0.84 for BlindDPS and 0.64/0.67/0.72 for KnownBlurDPS. Only the 90 mm value for BlindDPS (0.51) is indistinguishable from chance (within the 2.5th–97.5th percentile range of the label-permutation distribution, $p = 0.33$). BlindDPS performs better than KnownBlurDPS at 90 mm, the two methods are equal at 130 mm, and KnownBlurDPS performs better at 160 mm, which follows the CCC results (Fig. 5).

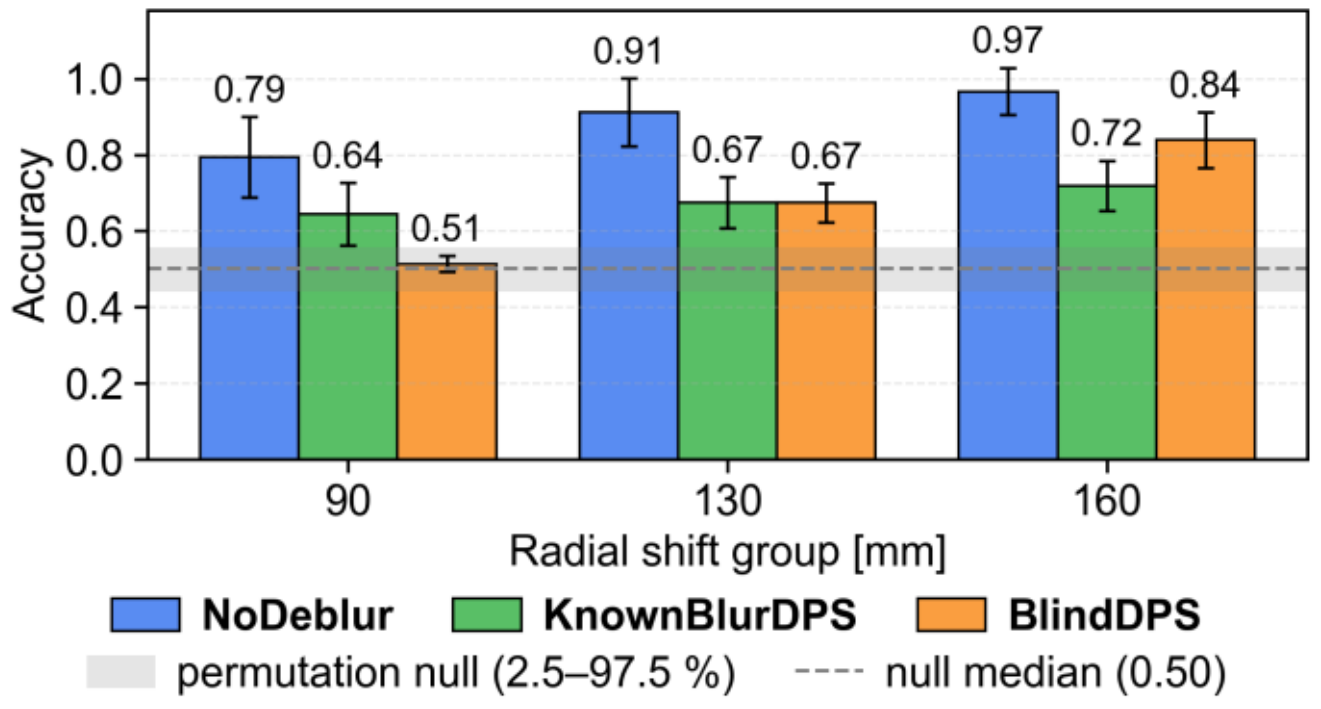

Fig. 8. Accuracy of a linear SVM separating the NoDeblur or DPS-restored patches of each shift group from the GT patches of the same locations (bars: mean over the five cross-validation folds; error bars: standard deviation over the folds). Dashed line: median of the label-permutation null (0.50); grey band: its 2.5th–97.5th percentile range.

## IV. Discussion

Both the known-kernel and blind DPS accomplished partial recovery of fine cancellous structures and image texture in trabecular bone patches distorted by shift-variant UHR CT system blur. Compared to original images acquired at 90–160 mm radial shift, both DPS variants resulted in significantly better fidelity and TF reproducibility to GT data acquired at isocenter (paired tests for SSIM/GMSD and paired permutation tests for CCC). The improvement in these performance metrics relative to the original images was particularly pronounced at 130 mm and 160 mm, likely because the resolution loss at these settings was appreciably worse compared to 90 mm shift. However, even at 90 mm, the number of reproducible TFs increased from 13 in NoDeblur to 33 with KnownBlurDPS and 40 with BlindDPS. Across all displacements, the DPS appeared to benefit TF reproducibility more than the IF scores – the increases in average TF concordance to isocenter were 30–130%, compared to the modest (although statistically significant) 1–10% gain in IF metrics. This observation reflects the fact that higher order grayscale value statistics captured by TFs are somewhat independent of the specific structures present in the image, so the texture may become more like that at the isocenter even if the morphology is not correctly resolved.

Similar to the comparisons between DPS and the unprocessed off-center image patches, the differences between the two DPS variants were also more pronounced in TF reproducibility than in IF metrics, with the blind approach yielding higher CCC at the moderate shifts (90 and 130 mm), whereas the known-kernel variant was more robust under the most severe blur (160 mm). The relative performance of the two methods at the different radial displacements can be attributed to two factors: 1) the known kernel for each shift was derived from MTF measurements normalized to a reference MTF at a 40 mm shift (the closest to isocenter datapoint provided in [1]), 2) the kernel estimation problem in BlindDPS is likely ambiguous, i.e., a narrower kernel with a blurrier restored image and a wider kernel with a sharper restored image can satisfy the same observed shifted trabecular patch. The first of these factors may explain why KnownBlurDPS could not restore images to better resolution characteristics than those of a 40 mm shift, leading to a possible under-sharpening relative to the GT acquired at <15 mm displacement. BlindDPS, on the other hand, could adapt the kernel to match the blur of GT data – the prior distribution was of sufficiently high resolution to support such estimates because the diffusion model had been trained on centered patches (<40 mm shift). This interpretation is consistent with the median blur width by BlindDPS being generally larger than the fixed, 40 mm-normalized known kernels, except for $\omega_u$ at 160 mm (Fig. 7). Conversely, it is likely that blind deblurring became unfavorable at this large shift (as observed in worse SSIM/GMSD and GLCM CCC compared to KnownBlurDPS) because of the ambiguity of kernel estimation in a setting where the measurement retained little high-frequency information to guide the optimization. As a result, BlindDPS under-estimated the radial blur at 160 mm relative to the MTF-derived value, while its azimuthal estimate was close to the MTF-derived one, yielding incomplete image feature recovery and possibly introducing a slight degree of unphysical anisotropy.

One of the primary objectives of this study was to assess if DPS can at least partly harmonize TFs of trabecular bone across different FOV locations to mitigate possible impact of patient positioning on the performance of downstream radiomic analysis. Our results indicate substantial but radial shift-dependent improvement in TF concordance to isocenter. While the reproducibility (CCC > 0.80) was almost completely restored at 90 mm, the results at the largest investigated radial displacement (160 mm) were more mixed, with 16 of the 40 features (including Cluster Shade, IMCs, MCC, Correlation, Autocorrelation, Joint/Sum Average and the high-gray-level run features) becoming reproducible with either DPS method. This compares favorably to only one feature (Cluster Shade) that was reproducible in the original 160 mm dataset, but leaves a number of TFs below 0.80 with both methods: two GLCM features (IDMN and MaxProb), most run-length features (LRE, SRE, RP, RLN, RLNN, RV), and the low-gray-level run features (LGLRE, LRLGLE, SRLGLE, which were only marginally improved at any shift). We conjecture that compared to TFs that became reproducible with DPS, these features are more sensitive to fine trabecular structures and low-intensity marrow background and therefore could not be harmonized at large shifts, where the morphological details of the patches were only partly restored by DPS despite their overall sharper appearance (as evidenced by SSIM and GMSD).

The classification study of Fig. 8 shows that without DPS, the radial shift of a bone patch can be detected from TFs with high mean accuracy – from ~80% at 90 mm to ~97% at 160 mm – underscoring the risk of confounding by patient position in radiomic prediction models. By improving TF reproducibility, both variants of DPS reduce shift classification accuracy across all displacements, although it remains >70% at 160 mm. Only at 90 mm with BlindDPS, the separability is reduced to chance level. These results are consistent with CCC analysis and suggest that DPS deblurring may be most advantageous for TF standardization within ~130 mm from isocenter. To put this finding in context, we note that patient centering errors of up to ~60 mm have been reported in clinical CT datasets, with most of the errors in the 10–30 mm range [27], [28]. For applications such as assessment of spine trabecular bone, we can expect additional ~50–100 mm variability in the positioning of a specific vertebral level across patients due to differences in body habitus and spine curvature. In this setting, most of the cases requiring TF harmonization would likely be within the range of moderate displacements where DPS (including BlindDPS) restores reproducibility for nearly all texture metrics.

To the best of our knowledge, our work is the first to report on the feasibility of DPS image restoration for mitigation of nonstationary system blur in x-ray CT. Most of the current literature on DPS in x-ray CT focuses on image reconstruction [12], [13], [29] using a known physical system model, in some cases including shift-invariant system or detector blur [13], [30]. In contrast, our approach acts on reconstructed images and demonstrates the possibility of estimating the system model in DPS jointly with the restoration, albeit for a relatively simple blur kernel (similar to the joint reconstruction-scatter correction of [31]). A similar post-reconstruction technique involving a diffusion prior paired with a likelihood term has been applied for low-dose denoising, but not deblurring [32]. Other diffusion-based methods for resolution recovery of CT images have been reported; however, in contrast to our work, they were trained on paired data [22], [33]. None of these studies addressed spatially-variant blur.

Another distinguishing aspect of this study is that our performance evaluations involve not only conventional image fidelity metrics but also assess the impact of DPS deblurring on texture image biomarkers. Compared to prior work on the application of diffusion models in improving TF reproducibility [16], which synthesized standardized CT images with a conditional latent diffusion model, our approach needs no paired training images. Because the radial shift enters only through the blur kernel, which BlindDPS estimates from the image itself, a prior trained once at isocenter can restore patches from any location in the FOV. These images can then be used for tasks beyond radiomic analysis. Further, while most of the efforts on TF harmonization in CT (regardless of the harmonization technique) attempt to address the effects of changes in scanner type [34], imaging protocols (e.g., imaging dose [35]), or reconstruction settings [16], [36], the proposed application of DPS to address loss of reproducibility due to changes in location within FOV (identified in [4]) appears unique.

Several limitations of this work need to be acknowledged. First, as mentioned earlier, the kernel estimates used in KnownBlurDPS were normalized to 40 mm radial shift instead of the ~15 mm displacement of the GT patches. However, the resulting underestimation of blur width was likely <10% (based on the trends in MTFs from [1]) and the current approach allowed us to avoid extrapolating below the range of available MTF measurements. Further, KnownBlurDPS may be impractical because of the difficulty of obtaining precise MTF estimates across the entire FOV. It was therefore included in this study primarily to establish performance bounds for evaluation of BlindDPS, which is likely better suited for clinical applications.

We also note that our evaluations involved rigid registration of the off-center scans of the test femur to its isocenter GT volume with resampling by linear interpolation. This step was necessary to enable paired IF and TF comparisons between image patches presenting identical underlying trabecular structure, but likely introduced slight blur in the images of the shifted bone. We believe that this was a minor effect because the estimated magnitudes of the rigid transformations were relatively small (e.g., rotation < 1°). However, it may have benefitted BlindDPS, which could inherently correct for this additional blur.

Another limitation of this work is that we only used normal femora for diffusion model training and DPS validation. We believe that this is acceptable for an initial feasibility study, but future work will need to investigate whether the diffusion prior needs to be enriched with pathological samples (e.g., osteoporotic or osteoarthritic bone) to maintain performance in general population. It should also be noted that we chose to perform our evaluations in 2D image slices to optimally use the available training data. The proposed approach can be extended to three-dimensional image restoration by stacking deblurred slices, although a native 3D diffusion model may ultimately be more desirable to ensure inter-slice consistency. Finally, our study was limited to UHR CT data from a single vendor and model (Canon Precision). However, the resolution properties of this device are representative of a broad category of emerging CT systems with fine detector pixels (~0.25 mm or less at the isocenter), including photon-counting CT. The system blur in all these UHR embodiments is likely to be more shift-variant than in conventional CT because it is dominated by source-side effects [1], motivating the need for the proposed restoration technique.

## V. Conclusion

We have demonstrated the feasibility of DPS deblurring in mitigating the effects of spatially-variant system resolution in UHR CT of trabecular bone. Compared to unprocessed scans obtained at 90–160 mm radial shift, DPS yielded statistically significant improvements in image fidelity and reproducibility of texture features to ground truth acquired at isocenter. Importantly, a blind variant of DPS achieved comparable performance to a known-kernel implementation, especially at moderate displacements of up to 130 mm. This has important practical implications, since it indicates that DPS deblurring can be deployed without extensive *a priori* system characterization. Following on our previous work that found a loss of reproducibility of trabecular bone TFs between locations in CT FOV due to spatially-variant blur, we have applied DPS for TF standardization. BlindDPS increased the mean CCC of GLCM and GLRLM TFs from 0.69 to 0.96 (GLCM) and from 0.65 to 0.90 (GLRLM) at 90 mm, and from 0.49 to 0.91 and from 0.44 to 0.84 at 130 mm; the classification accuracy for distinguishing bone patches acquired at 90 mm from those at isocenter was reduced from ~80% to chance level with BlindDPS. These results suggest that DPS deblurring should be considered for TF harmonization against the effects of patient mis-centering in UHR CT and motivate future work focused on evaluating this approach in radiomic models for specific clinical applications.